**Effect of annealing on the hot salt corrosion resistance of the fine-grained titanium α-alloy Ti–2.5Al–2.6Zr obtained via cold Rotary Swaging**

V.N. Chuvil'deev [a], A.V. Nokhrin[a]*, C.V. Likhnitskii [a], A.A. Murashov [a], N.V. Melekhin [a], K.A. Rubtsova[a], A.M. Bakmetyev[b], P.V. Tryaev[b], R.A. Vlasov[b], N.Yu. Tabachkova[c, d], A.I. Malkin[e]

[a] Lobachevsky State University of Nizhniy Novgorod, 23 Gagarin Ave., Nizhny Novgorod, 603022, Russia

[b] Afrikantov Experimental Design Bureau for Mechanical Engineering, JSC, 15 Burnakovskiy Lane, Nizhniy Novgorod, 603074, Russia

[c] National University of Science and Technology "MISIS", 4 Leninskiy Ave., Moscow, 119991, Russia

[d] A.M. Prokhorov General Physics Institute of the Russian Academy of Sciences, 38 Vavilov st., Moscow, 119991, Russia

[e] A.N. Frumkin Institute of Physical Chemistry and Electrochemistry of the Russian Academy of Sciences, 31 Leninskiy ave., Moscow, 119071, Russia

E-mail: chuvildeev@nifti.unn.ru

**Abstract**

A hot salt corrosion (HSC) test was performed on the fine-grained titanium α-alloy Ti–2.5Al–2.6Zr (Russian industrial alloy PT–7M). The ultrafine-grained (UFG) microstructure in the titanium α-alloy was formed via cold Rotary Swaging. The grain size and volume fraction of the recrystallized microstructure in the alloy were varied by choosing appropriate annealing temperatures and times. The microstructure and corrosion resistance of UFG alloys were studied after 30 min of annealing at 500–700°C and after 1000 h of annealing at 250°C. Metallographic studies were carried out to investigate the effects of annealing on the nature and extent of corrosive damage in the titanium α-alloy Ti–2.5Al–2.6Zr. After HSC tests, surface analyses of the titanium α-alloy samples were conducted using X-ray diffraction and electron microscopy. During the HSC testing of the titanium α-alloy Ti–2.5Al–2.6Zr, a competitive interaction between intergranular corrosion (IGC) and pitting corrosion was observed. To the best of our knowledge, it was shown for the first time that annealing affects the relationship among the IGC, pitting corrosion and uniform corrosion rates of the titanium alloy. Prolonged low-temperature annealing at 250°C resulted in a more pronounced increase in the uniform corrosion rate than short-term high-temperature annealing for 30 min at 500-700°C. An in-depth analysis of the effect of the structure and phase composition of the grain boundaries on the susceptibility of the α-alloy Ti–2.5Al–2.6Zr to HSC was conducted.

---

* Corresponding author (nokhrin@nifti.unn.ru)

**Abbreviations:** ECAP – equal channel angular pressing; HAGBs – high-angle grain boundaries; HSC – hot salt corrosion; IGC – intergranular corrosion; RS – rotary swaging; RT – room temperature; SEM – scanning electron microscopy; TEM – transmission electron microscopy; UFG – Ultrafine-grained (materials); XRD – X-ray diffraction (analysis).

## 1. Introduction

Titanium α- and near-α alloys with a low content of β-phase particles are widely used in manufacturing the corrosion-resistant structural components in nuclear power engineering and marine aviation. For instance, α- and near-α titanium alloys, such as Ti–Al–(V, Zr), are used in the production of heat-exchange equipment for nuclear power plants [1-3], while two-phase titanium alloys are employed in aviation [4] and marine engineering [3, 5]. Modern titanium α-alloys are subject to stringent requirements regarding strength, resistance to hot salt corrosion (HSC), corrosion fatigue, stress corrosion cracking and hydrogen embrittlement.

HSC is one of the most critical degradation processes for titanium alloys in nuclear power engineering and marine aviation [1, 3, 6, 7]. This phenomenon involves the formation of thick salt films on the surfaces of titanium alloy products, causing accelerated corrosion at elevated temperatures [6, 7]. Intergranular HSC, occurring under conditions of free access to oxygen and the presence of water vapour, is especially hazardous [8]. Intergranular corrosion (IGC) leads to the formation of micro-crack-type defects starting from the grain boundaries. These grain boundary micro-cracks will cause accelerated fracture in high-performance titanium structures operating under static and cyclic mechanical stresses. HSC is especially hazardous when titanium alloys are used under static or cyclic stresses, thereby accelerating hydrogen embrittlement and hot salt stress corrosion cracking [6-13]. Therefore, the study of HSC mechanisms in titanium alloys is required.

Typically, the strength of titanium alloys increases with increasing volume fraction of the β-phase [4, 5]. However, this approach is inefficient for radiation-resistant titanium alloys used in nuclear power engineering. This issue is often addressed by selecting additional alloying elements, including the platinum group metals [2, 14, 15], which significantly increases the cost of titanium alloy semi-finished products and manufactured items. Therefore, alternative methods for enhancing the operational characteristics of titanium alloys are being extensively developed, including new manufacturing technologies, the titanium alloy–composition optimisation methods [16-18] and surface modification methods, which allow improving the corrosion resistance of titanium alloys [19-21].

Developing the ultrafine-grained (UFG) microstructures is an efficient method for improving the mechanical properties and operational characteristics of titanium alloys [22, 23]. UFG titanium alloys exhibit enhanced strength [22, 24], increased fatigue strength [22, 25, 26] and superplasticity at elevated temperatures [27, 28] as well as improved corrosion resistance [29, 30] and resistance to stress corrosion cracking in some cases [31]. In the future, the high strength and good corrosion resistance of the UFG titanium alloys will enable the manufacturing of lighter and more reliable structures.

The resistance of UFG titanium α-alloys to HSC has been scarcely investigated. In [32], Equal Channel Angular Pressing (ECAP) at elevated temperatures (450°C) was demonstrated to increase the HSC resistance of the Ti–5Al–2V near-α alloy by reducing the local vanadium concentration at the grain boundaries. The grain boundary migration process during HSC led to an increased local vanadium concentration at the grain boundaries, contributing to accelerated IGC. High-speed solid-phase diffusion welding, which reduces the grain boundary migration rate during heating, allows the production of welded products with increased hardness and corrosion resistance from the near-α alloys with a UFG microstructure [33]. The corrosion resistance of the Ti–2.5Al–2.6Zr α-alloy obtained through Rotary Swaging (RS) was investigated for the first time in [34]. The grain boundary migration was shown to begin during hot-corrosion testing owing to the low thermal stability of the non-equilibrium UFG microstructure, which can lead to changes in the HSC mechanism. This finding is unexpected because traditionally, the mechanism of corrosion damage is assumed to remain the same during testing. The hypothesis of the invariability corrosion damage mechanism allows predicting the service life of products used in corrosive environments for extended periods.

The present study aims to extend the research conducted in [34], focusing on exploring the impact of grain boundary migration and the volume fraction of recrystallized microstructures on the susceptibility to HSC of the titanium α-alloy Ti–2.5Al–2.6Zr, which is manufactured using cold RS. RS technology has emerged as a promising approach for creating gradient UFG microstructures in alloys that are difficult to deform [35-37], sparking considerable interest among researchers in RS as an efficient and cost-effective method for producing high-strength, functionally graded materials such as titanium [34-37], zirconium [38] and copper alloys [39], as well as stainless steel [40]. Recent studies have shown significant potential for RS in enhancing the strength of titanium while preserving its good ductility [41]. This is accomplished by implementing multiple non-uniform deformations during RS, which result in the developing of a unique non-uniform microstructure that contributes to further strength enhancement [41-46]. The issue of whether high strength and HSC resistance in titanium alloys can be concurrently attained through the creation of a non-uniform microstructure remains unexplored (refer to reviews [42, 43]).

## 2. Materials and methods

The study focused on Russian industrial titanium α-alloy grade PT-7M with the composition Ti–2.45 wt.% Al–2.63 wt.% Zr (refer to Table 1). The alloy composition meets the requirements of Russian National Standard GOST 19807-91. Rods of PT-7M alloy, measuring 20 mm in diameter, were produced at the Chepetsk Mechanical Plant, JSC (Russia, Glazov). PT-7M alloy tubes are employed under conditions conducive to HSC and stress corrosion cracking, namely at temperatures up to 250°C and in the presence of salt deposits of chlorides and bromides on the tube surfaces.

**Table 1.** Composition of titanium alloy (by wt.%)

| Ti | Al | Zr | Si | Fe | V | Nb | $O_2$ | $H_2$ | $N_2$ | C |
|---|---|---|---|---|---|---|---|---|---|---|
| Balance | 2.45 | 2.63 | 0.02 | 0.086 | 0.002 | 0.024 | 0.12 | 0.001 | 0.003 | 0.028 |

The microstructure of the titanium alloy Ti–2.5Al–2.6Zr was developed through RS at room temperature (23–25°C). Rods of PT-7M alloy with an initial diameter of 20 mm were processed using the RS5-4-21 HIP machine. The rods were deformed using four dies made of high-strength steel. RS of the titanium rods reduced their diameter from the initial value of 20 mm to 12 mm in 2-mm decrements (⌀20 mm → 18 mm → 16 mm → 14 mm → 12 mm) followed by 1-mm decrements (⌀12 mm → 11 mm → … → 7 mm → 6 mm). The final diameter of the titanium rods after RS was 6 mm, corresponding to a total strain of 70%. The strain rate was measured at 0.6–1 s$^{-1}$. The rotation speed of the workpiece was approximately 60 rpm, with a hammer strike frequency of 1200 strikes per minute and a workpiece feeding rate of 50 mm/min. The temperature of the outer surfaces of the titanium rods did not exceed 60°C–100°C during RS.

The samples were subjected to annealing under two regimes. Regime I involved 30-min annealing of titanium samples in the air at 500°C–700°C. Regime II involved annealing at 250°C for durations of 500 and 1000 h. The samples were cooled under ambient air. Annealing under

Regime I was used to change the volume fraction of the recrystallized microstructure and the grain sizes of the titanium α-alloy. Annealing under Regime II was designed to assess the long-term thermal stability of the UFG structure of the alloy.

The HSC tests were conducted in NaCl at 250°C with free access to air. The temperature was maintained with an uncertainty of ±5°C. The salt contained impurities of K, Ca, Mg and S (Fig. 1). The HSC tests were conducted for 250 h in laboratory autoclaves with a volume of 3000 cm$^3$. For the tests, samples were placed in the centre of ceramic beakers with the internal volume of 600 cm$^3$ that allowed air access. These beakers were filled with crystalline salt. The ceramic beakers containing the titanium alloy samples immersed in the salt were placed in the centre of the test autoclaves. The HSC testing methodology is described in detail in [47]. Cylindrical samples with a diameter of 5 mm and a length of 20 mm were used for the tests. To determine the corrosion depth, cross–sections of the samples were prepared. The corrosion depth was determined using a Leica IM DRM optical microscope according to the requirements of the Russian National Standard GOST 9.908-85. The average uncertainty in determining the corrosion defect depth was ±10 μm.

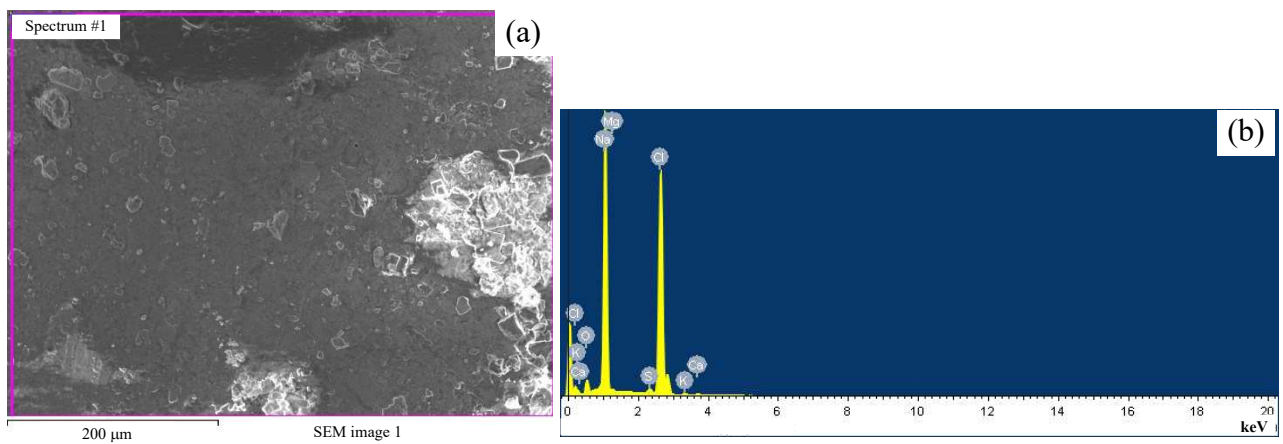

**Fig. 1**. SEM image (a) and EDS microanalysis results (b) of the NaCl salt used in the HSC testing.

The sample microstructure was investigated using scanning electron microscopy (SEM), Tescan Veg 2 and JSM-6490 equipped with an Oxford Instruments INCA 350 EDS microanalyzer, and transmission electron microscopy (TEM) JEM-2100F. The phase analysis of the crystalline corrosion films was conducted via X-ray diffraction (XRD) using a Haoyuan DX27 diffractometer (CuK$_α$ emission, scanning step = 0.04°, exposure time at every step = 2 s and scanning angle range

2θ = 30–80°). The XRD analysis of the salt-film phase compositions after HSC testing could not be conducted owing to the high NaCl phase content, which resulted in excessively intense XRD peaks. Therefore, before conducting the XRD experiments, the salt films were mechanically removed from the sample surfaces and subsequently rinsed with hot distilled water using filter paper (Fig. 2a). The insoluble residue left after washing was air dried for 10 min and finely ground in an agate mortar to obtain a homogeneous powder. This method had been used previously to assess the composition of salt films after HSC testing of titanium alloys [16]. The salt samples were placed in an aluminium trough so that the area exposed to X-rays closely matched the area marked by the white rectangle in Fig. 2b. This arrangement aimed to maximise the sample area irradiated during the experiment. Despite the efforts, not all of the NaCl phase was removed from the corrosion films of titanium alloys. As a result, highly intense peaks of the NaCl phase were observed in the XRD curves of most of the studied samples. However, a considerable reduction in the intensities of NaCl peaks compared with those of the original samples allowed qualitative analysis of their phase composition. Qualitative phase analysis was performed using the DIFFRAC.EVA software, and the results were analysed using the PDF-2 (ICDD, release 2012), PDF-4 (ICDD, release 2014) and ICSD (2016) databases.

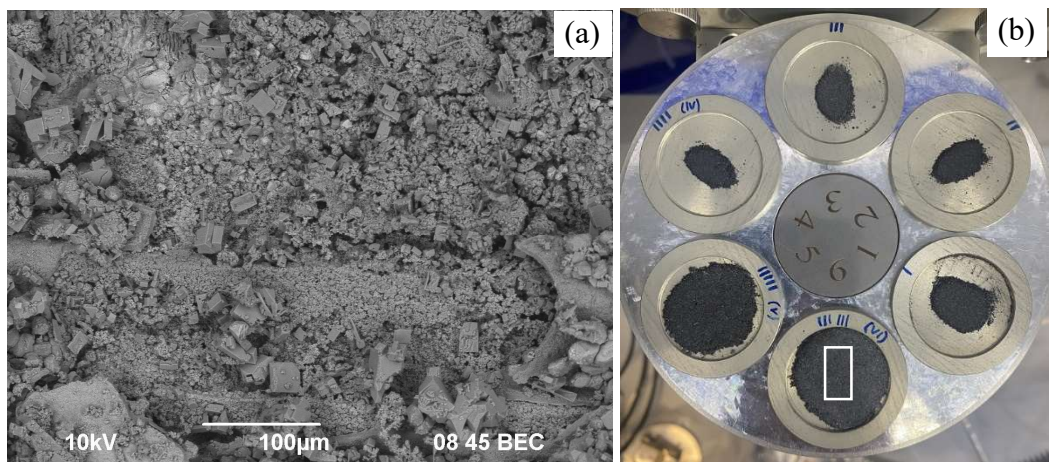

**Fig. 2**. (a) SEM–image of the salt residue on the surfaces of the Ti–2.5Al–2.6Zr alloy samples and (b) method for positioning the salt-film samples studied in the aluminium trough. The white rectangle indicates the approximate area irradiated by the X-ray beam.

The microhardness (Hv) measurements were performed using a Qness 60A+ hardness tester. The compression stress relaxation tests were carried out using the samples of 5-mm diameter and 10-mm height according to the procedure described in [48], in accordance with the requirements of Russian National Standard GOST R 57173-2016. During the stress relaxation tests, the macro-elastic stress ($\sigma_0$) and the physical yield strength ($\sigma_y$) were determined. According to [48], the macro-elastic stress values indicate the material crystal lattice yielding threshold stress.

## 3. Results

### 3.1 Microstructure and mechanical properties

The results of the microstructure investigation of the Ti–2.5Al–2.6Zr alloy after cold RS have been partially described in [34, 49]; therefore, we will not reiterate them extensively here. The data presented in [34, 49] will be complemented with new experimental findings necessary for analysing the results.

In the initial state, the PT-7M alloy exhibited a coarse-grained microstructure consisting of a small number of equiaxial α-grains and α′-phase plates (Fig. 3a). This microstructure is typical for the PT-7M alloy obtained via hot deformation with gradual temperature reduction [50-52]. The thickness of the titanium α′-phase plates initially ranged from 5 to 10 μm, with their length reaching several hundred μm. The EBSD analysis of the microstructure showed that the majority of α′-phase plates in the coarse-grained Ti–2.5Al–2.6Zr alloy were surrounded by low-angle boundaries (Fig. 3b, indicated in white). A fine precipitation of β-phase micro-particles was observed along the grain boundaries of the titanium α′ phase (Fig. 3c). The quantity of β-phase particles located along the boundaries of the equiaxial α-grains was very small (~0.4 vol.%) (Fig. 3d).

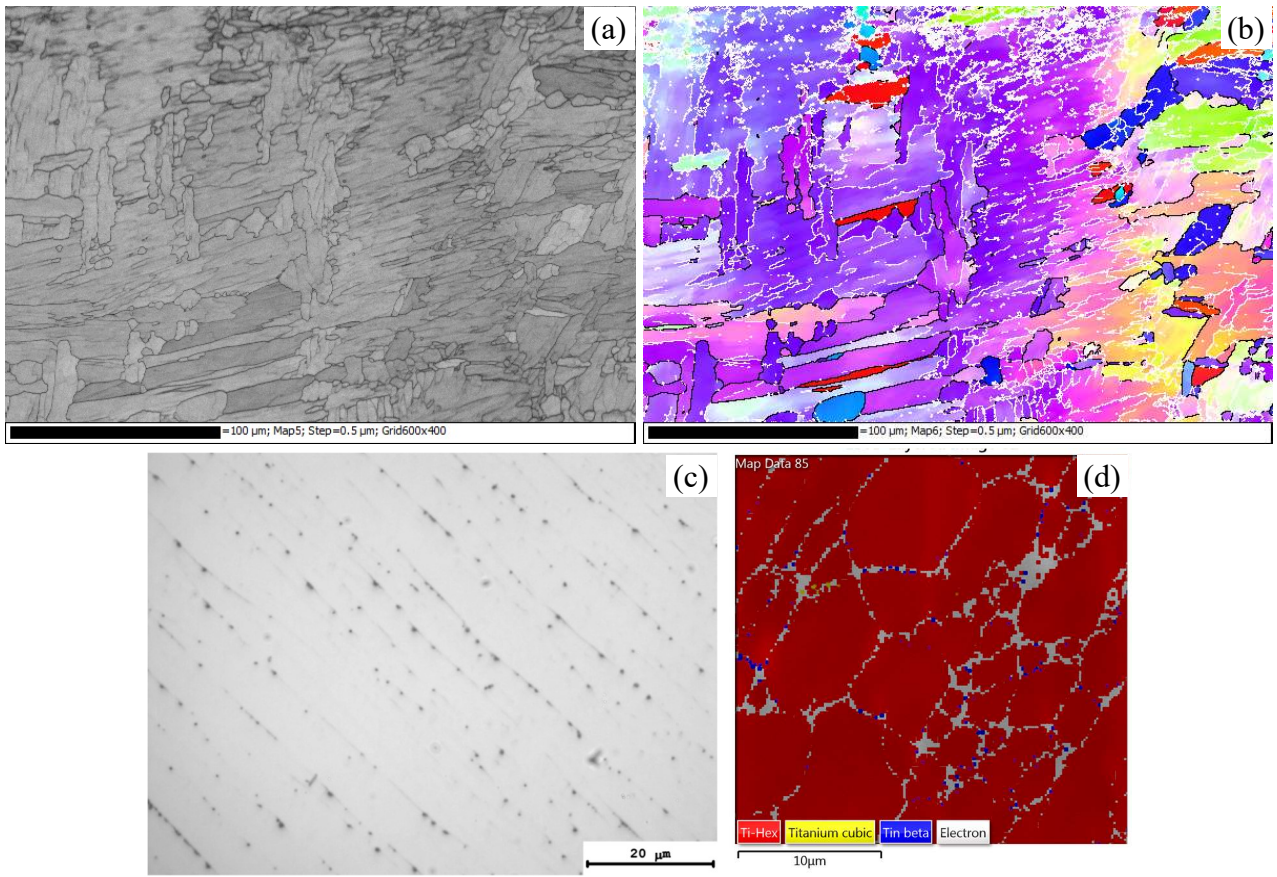

**Fig. 3**. Microstructure of the Ti–2.5Al–2.6Zr alloy in its initial state: (a) a general view of the microstructure; (b) EBSD microstructure analysis, (c) β-phase particles located along the boundaries of the α′-phase plates and (d) EBSD analysis focusing on the equiaxial α grains.

The RS process induces the formation of a non-uniform macrostructure within the titanium alloy. As the titanium alloy workpiece undergoes rotation and elongation during the RS process, a non-uniform vortex-like macrostructure emerges in the cross-section of the rod (Fig. 4a) and a striated fibrous-type macro-structure appears in the longitudinal section (Fig. 4b).

After RS, the coarse-grained structure develops into a fragmented microstructure, with an average fragment size of 0.5–1 μm (Fig. 4c). The electron diffraction patterns show good correspondence to α-Ti. EDS micro-analysis results reveal the grain boundaries of the Ti–2.5Al–2.6Zr alloy to be enriched with Zr; the local Zr concentration within the grains was found to be 1.6–2.0%, whereas at the fragment boundaries, it increased to 3.1-3.6% (Fig. 4d). It is noteworthy that the elevated Zr concentration at the grain boundaries in the Ti–2.5Al–2.6Zr alloy (spectra #8 and

#10 in Fig. 4e) induces a competing segregation effect. Consequently, the Al concentration at the grain boundaries in the Ti–2.5Al–2.6Zr alloy, 1.4–1.8%, is lower than that within the grains (2.1–2.3%). The Zr concentration at the low-angle grain boundaries matches that inside the grains (spectrum #11 in Fig. 4e).

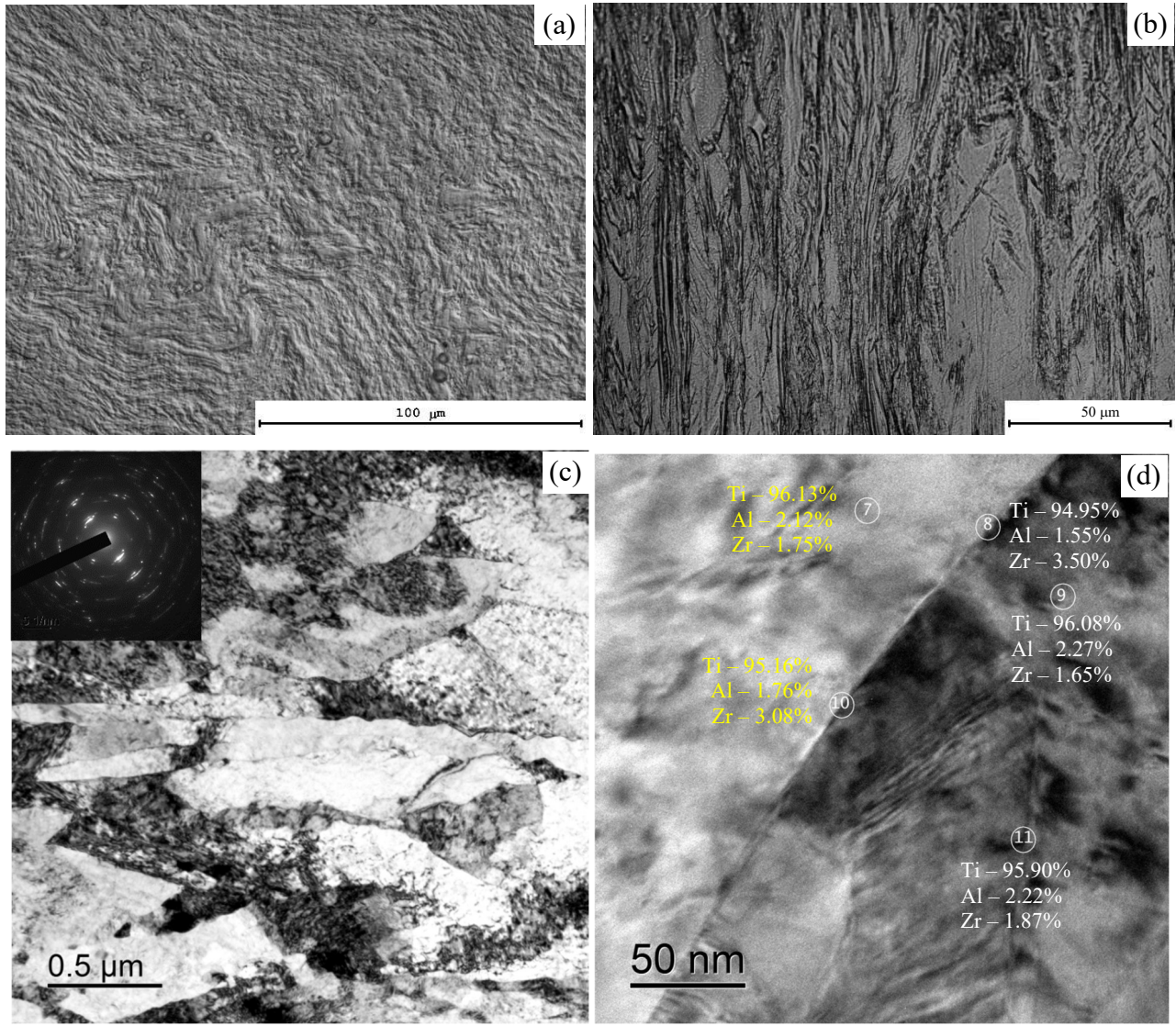

**Fig. 4**. Macro- and microstructure of the UFG alloy Ti–2.5Al–2.6Zr following RS: (a, b) microstructure of the cross-sectional (a) and longitudinal sectional (b) views of the central part of the rod, (c) microstructure of the UFG alloy post-RS [34, 49] and (d) compositional analysis of the grain boundaries in the UFG alloy after RS [49].

The hardness (Hv) of the coarse-grained Ti–2.5Al–2.6Zr alloy in its initial state was approximately 2.0 GPa. The hardness of the fine-grained alloy decreases from 2.9–3.0 GPa down to

2.5 GPa from the edge to the centre of the sample cross section (Fig. 5a). The non-uniform nature of the titanium macrostructure after RS is clearly visible on the surface of the etched specimen (Figs. 5b and c). The tension tests revealed an increase in the tensile strength ($\sigma_b$) of the Ti–2.5Al–2.6Zr titanium alloy from 590 up to 1080 MPa after RS [49]. At the same time, the relative elongation to failure ($\delta_5$) decreased from 40% down to 6–8% [49].

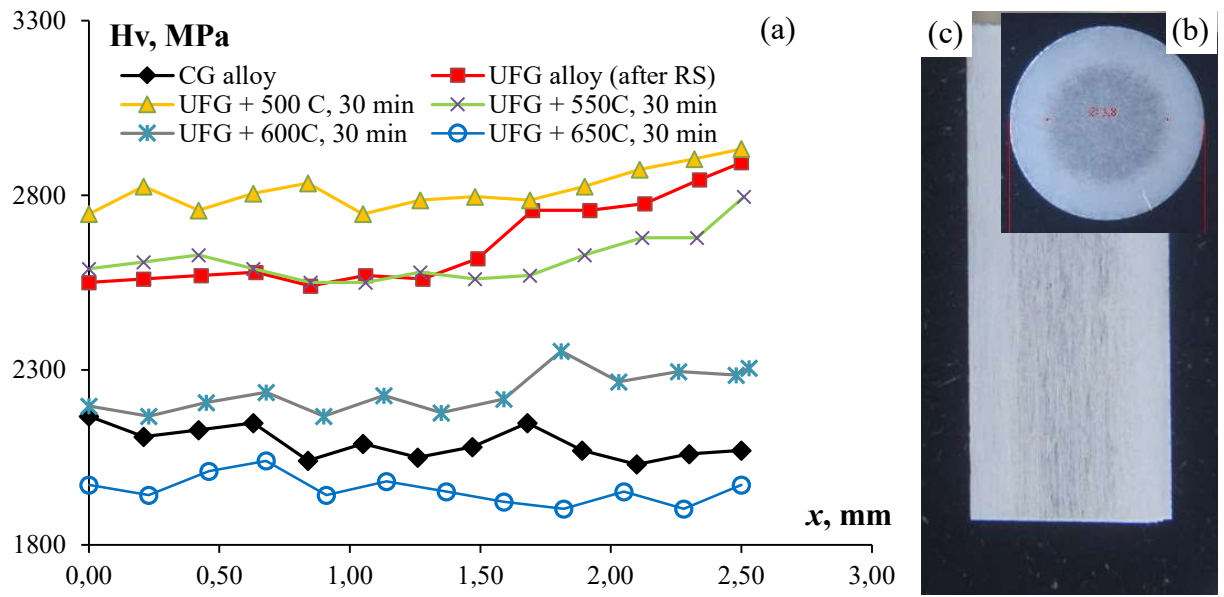

**Fig. 5.** Hardness distribution across the cross section of a Ti–2.5Al–2.6Zr alloy sample post-RS and following various annealing processes. Fig. 5b and 5c illustrate the etched macrostructure of the longitudinal (b) and cross-sectional (c) views of the sample after RS.

Prolonged isothermal annealing at 250°C results in an increase in the microhardness of the Ti–2.5Al–2.6Zr alloy. After 1000 h of annealing at 250°C, the microhardness of the central part of the rod experienced a slight increase from 2.5 GPa to 2.65 GPa.

Annealing the Ti–2.5Al–2.6Zr alloy at 400°C for 30 min led to the precipitation of Zr particles and initiated recovery processes, accompanied by a decrease in the density of lattice dislocations as the annealing temperature increased. The size of the precipitated Zr particles ranged from 20 to 50 nm (Figs. 6a and 6b). Upon heating up to 400°C, nano-metre-sized TiFe$_2$ particles were detected in the microstructure of the titanium alloy (Fig. 6c), along with TiFe particles (Fig.

6d). The potential formation of nano-meter-sized Ti–Fe intermetallic particles in the Ti–Al–Zr alloy was demonstrated in [53].

After annealing at 500–550°C for 30 min, the recrystallization processes commenced in the Ti–2.5Al–2.6Zr titanium alloy, leading to the dissolution of Zr particles. The sizes of the Ti–Fe particles remained largely unchanged. Within the recrystallized grains of the Ti–2.5Al–2.6Zr titanium alloy, plate-like particles, presumably of the α″-phase, formed (Figs. 6c, 6d and 7). The formation of the α″-phase after annealing the near-α alloy was also demonstrated in [54]. Following annealing at 600-700°C for 30 min, the density of lattice dislocations significantly decreased, and Zr particles were absent in the microstructure of the recrystallized alloy.

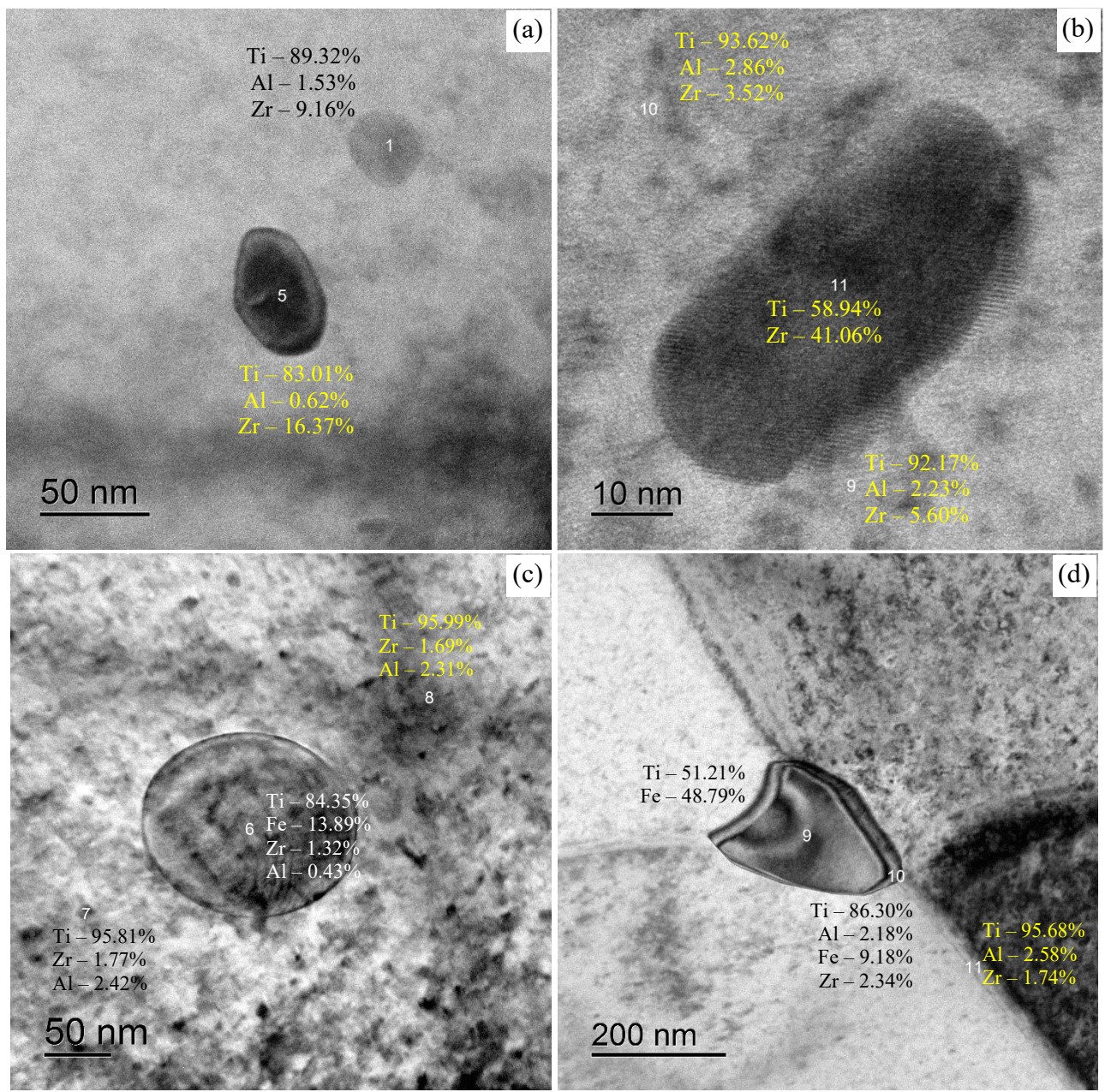

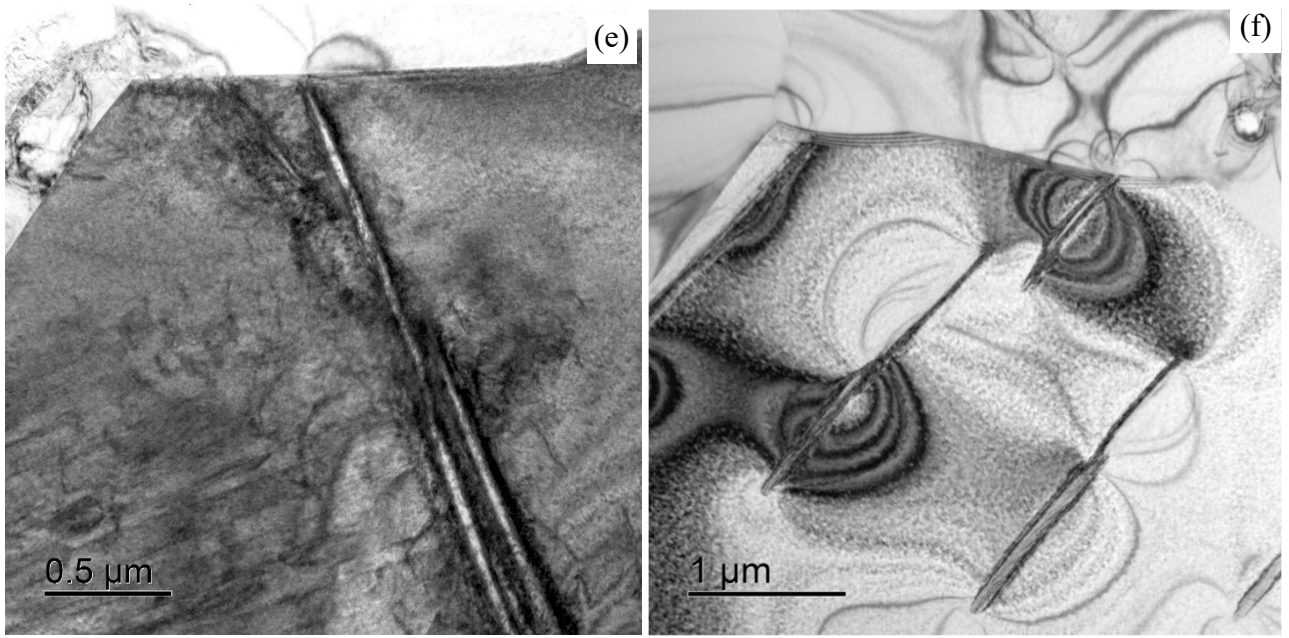

**Fig. 6.** Nucleation of particles in the microstructure of annealed UFG alloy Ti–2.5Al–2.6Zr: (a, b) Zr particles observed after annealing at 400°C (a) and 500°C (b) [49]; (c, d) TiFe$_2$ (c) and TiFe (d) particles noted after annealing at 400°C; (e, f) α″-phase particles within recrystallized α-Ti grains after annealing at 600°C (e) and 700°C (f) [49].

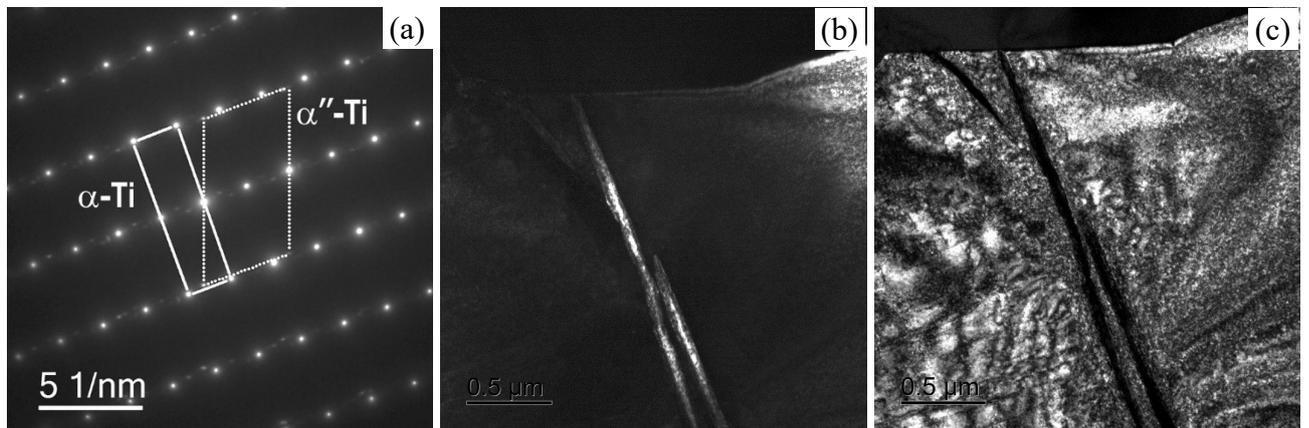

**Fig. 7.** (a) Analysis of the α″-phase particles in recrystallized α-Ti grains after annealing at 500°C. Dark-field images in the (b) α-phase and (c) α″-phase reflections. TEM [49]

The precipitation of Zr particles and the α″-phase within the grains of the Ti–2.5Al–2.6Zr titanium alloy results in an increase in macro-elastic stress ($\sigma_0$) at 400–450°C and 600°C, respectively (Fig. 8). The reduction in macro-elastic stress of the Ti–2.5Al–2.6Zr alloy at higher annealing temperatures is primarily attributed to a decrease in the density of lattice dislocations.

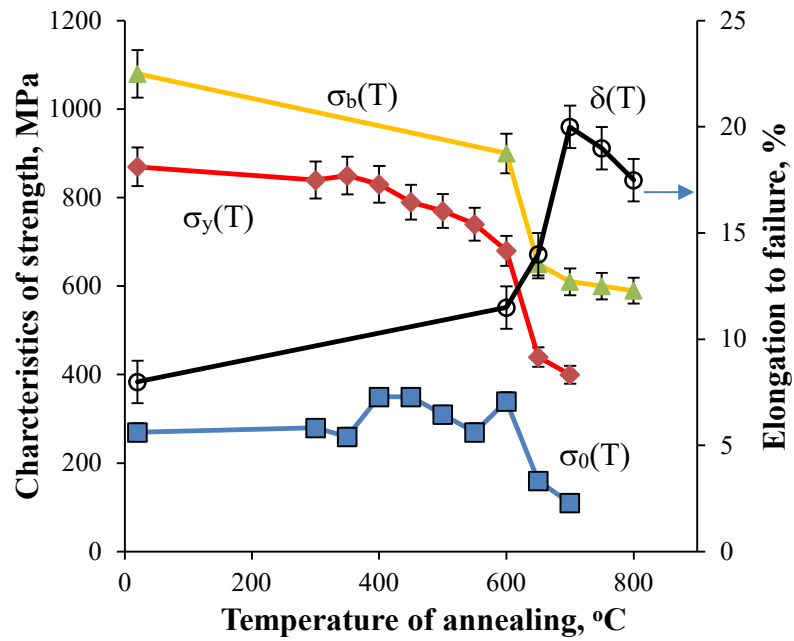

**Fig. 8.** Dependencies of the macro-elastic stress ($\sigma_0$), yield strength ($\sigma_y$), tensile strength ($\sigma_b$) and elongation to failure ($\delta$) on the annealing temperature of the UFG alloy Ti–2.5Al–2.6Zr.

At the initial stage of recrystallization annealing, an unusual multi-grained structure was formed in the fine-grained titanium alloy Ti–2.5Al–2.6Zr (Figs. 9a and 9b). After annealing at 650°C and 700°C, a uniform fine-grained microstructure was formed in the alloy (Figs. 9c and 9d). After annealing at 650°C, the average grain size ranged from 2.8–3.0 μm (in the surface layer) to 3.5–4 μm (in the central part of the rod). The difference in hardness between the central and surface zones in the cross-section of the rod did not exceed 0.1–0.15 GPa (Fig. 5). The mechanical properties of the annealed Ti–2.5Al–2.6Zr alloy are as follows: tensile strength $\sigma_b$ = 590–620 MPa and elongation to failure $\delta_5$ = 17.5–20% (Fig. 5).

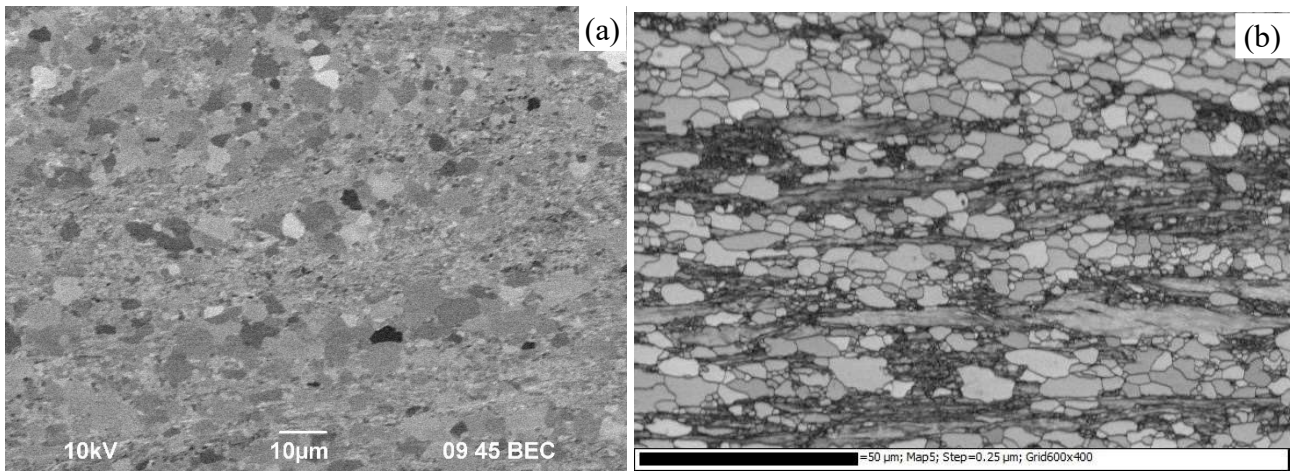

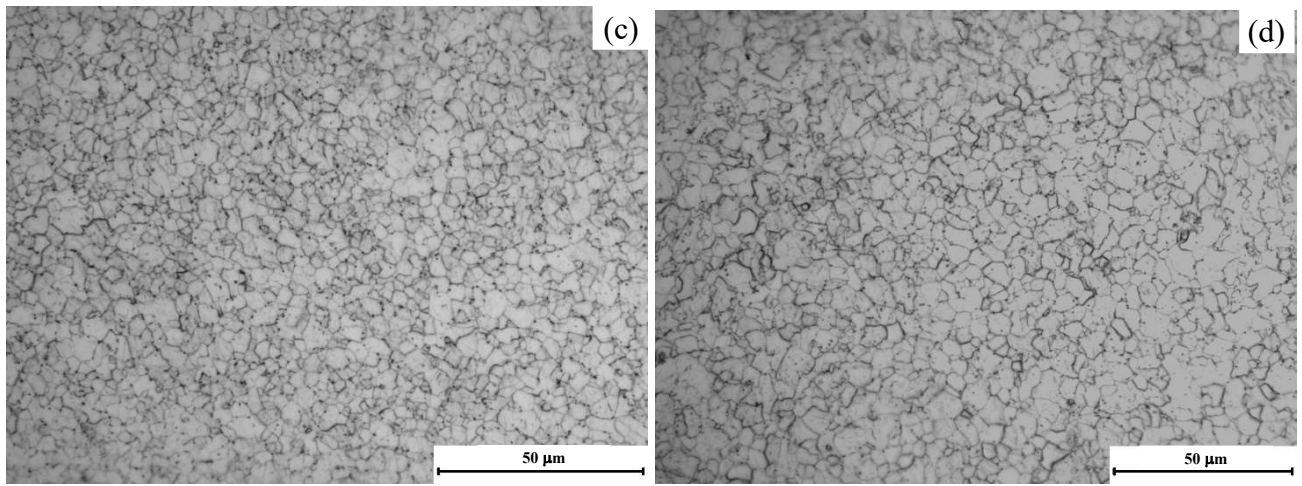

**Fig. 9**. Microstructure of the (a, c) surface layers and (b, d) central parts of the Ti–2.5Al–2.6Zr alloy rods after RS and annealing at (a, b) 600°C and (c, d) 650°C [49].

During the recrystallization annealing, alongside the grain growth, there was a decrease in the fraction of low-angle boundaries (Fig. 10a). In the recrystallized titanium alloy, the high-angle grain boundaries (HAGBs) constituted the majority of boundaries (Fig. 10b).

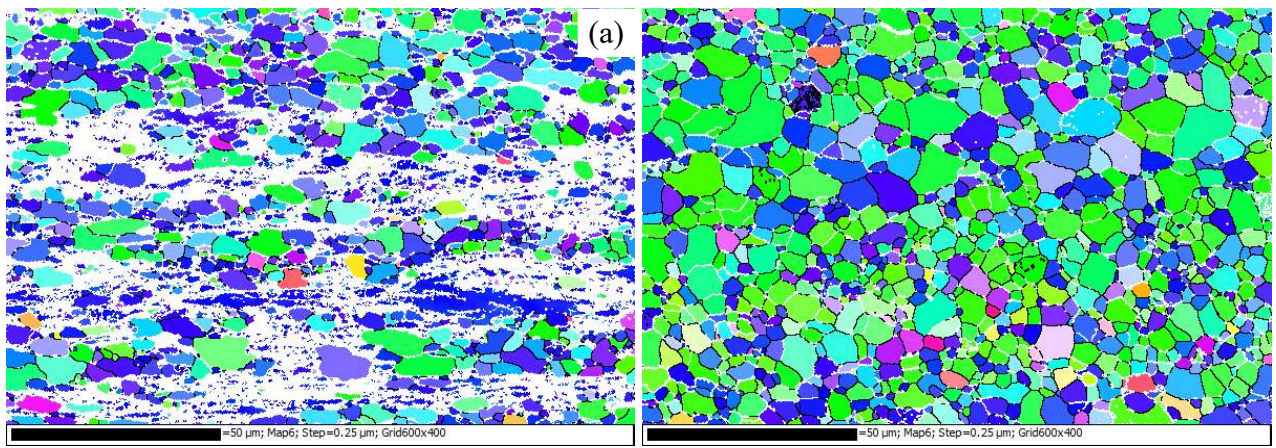

**Fig. 10** EBSD analysis of the grain boundary misorientation spectra in the Ti–2.5Al–2.6Zr alloy: (a) UFG alloy after annealing at 600°C and (b) UFG alloy after annealing at 650. SEM.

3.2 Corrosion tests

The residues of NaCl salt used in the corrosion testing were present on the surface of the titanium alloy samples after HSC testing (Fig. 11a). After rinsing with a stream of hot water, dark films firmly bonded to the sample surfaces were observed on the samples (Fig. 11b). It was not possible to mechanically remove these films without damaging the sample surfaces. After chemical cleaning and removal of these films, large corrosion pits were found on the surfaces of all samples (Fig. 11c). The corrosion pits were uniformly distributed across the sample surfaces, indicating that the salt particles adhered uniformly to the sample surfaces during corrosion testing.

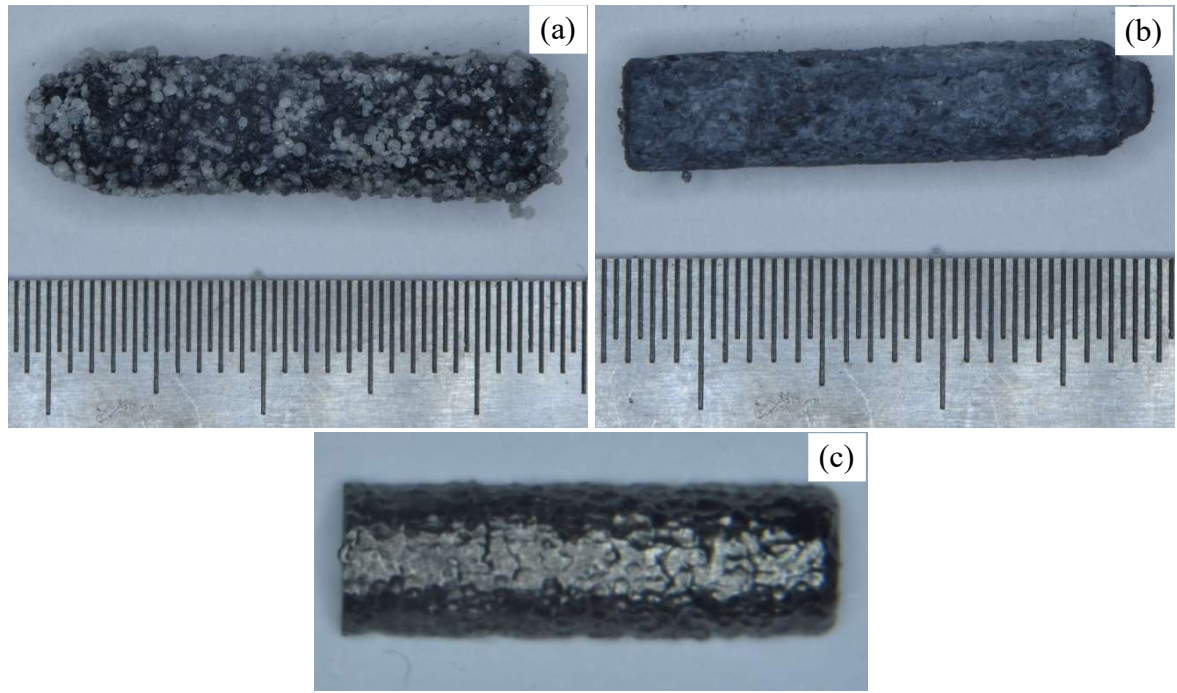

**Fig. 11.** Typical appearance of the titanium alloy samples after the HSC tests: (a) after the HSC test, (b) after washing in hot water and (c) after chemical cleaning according to GOST R 9.907-2007.

The XRD analysis results of the corrosion of the powdered products placed into the Al trough (Fig. 2) are presented in Fig. 12. The XRD spectra of the corrosion films from the coarse-grained alloy reveal highly intense peaks corresponding to the NaCl phase (PDF #00-005-0628) and the Al trough (Fig. 12a) as well as low-intensity peaks of rutile $TiO_2$ (PDF #00-021-1276), which constitutes the main phase, along with mon-TiO (PDF #01-072-0020) and $Ti_2O_3$ (PDF #00-010-0063). Additionally, the XRD spectra of the corrosion films indicate the presence of α-Ti with two sets of lattice parameters. The first set (a, c) corresponds to card PDF #00-044-1294 (ICSD #44390), and the second set (a, c) exhibits an increased unit cell volume compared with card PDF #00-044-1294 (ICSD #44390). The peak corresponding to metallic Ti in the XRD results is likely attributed to the stage of mechanical separation of the corrosion films from the surface of the tested samples (see Materials and Methods section). The HSC films of the UFG alloy comprise the same phases as those found in a coarse-grained alloy (α-Ti, rutile TiO$_2$, mon-TiO and Ti$_2$O$_3$) and possibly the Ti$_3$Al phase (PDF #00-052-0859) (Fig. 12b).

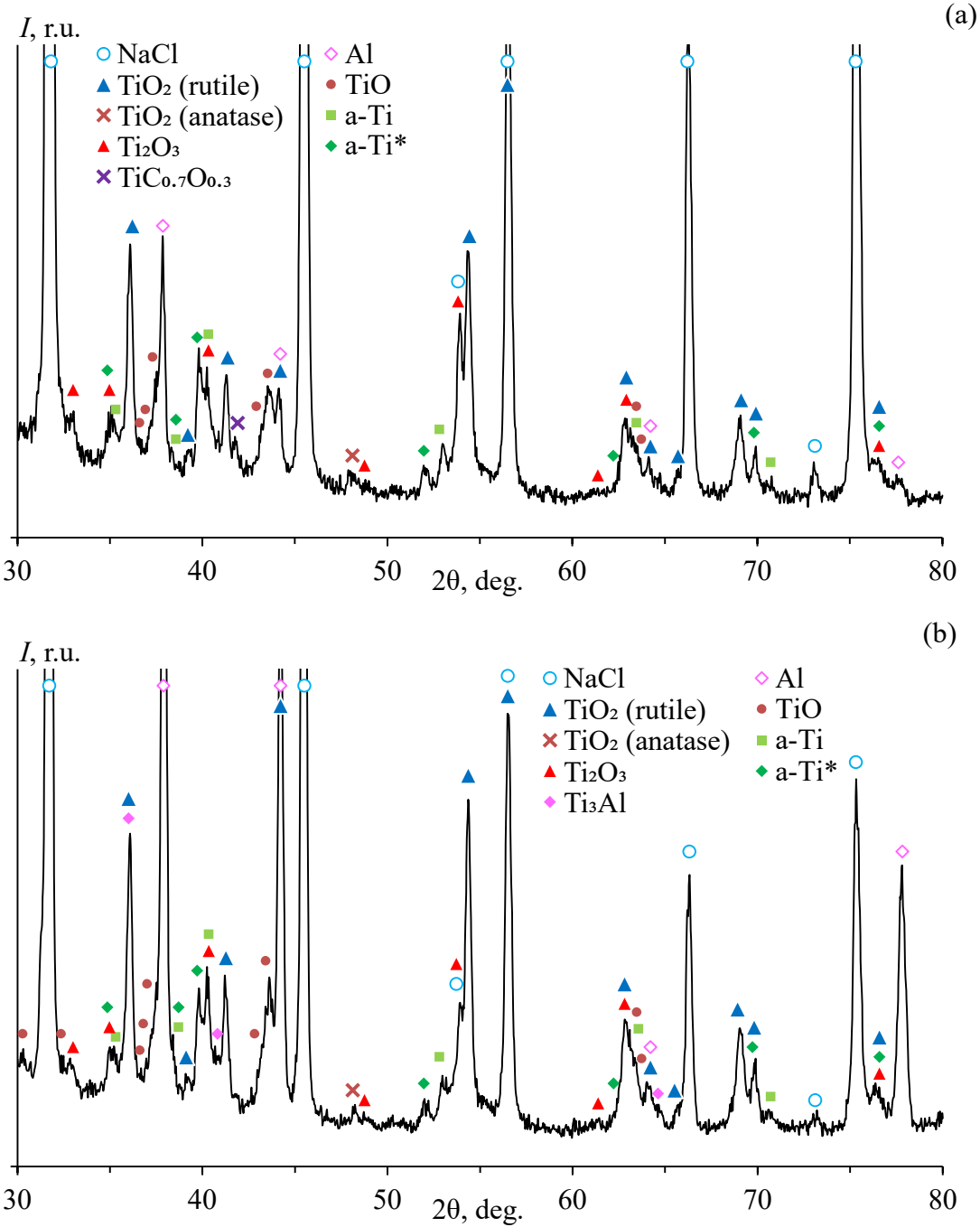

**Fig. 12**. Results of XRD investigations of the salt films of (a) coarse-grained and (b) UFG alloys.

The EDS micro-analysis revealed that the salt deposits on the surfaces of the coarse-grained samples (Fig. 11a) contained Ti, Cl, Na, Al and oxygen (O) (Fig. 13). The presence of sulfur (S) in the corrosion films was associated with the presence of S in the NaCl salt used for the HSC testing

(Fig. 1). The composition of the salt deposits on the surfaces of the UFG alloy samples was found to be similar to that of the salt on the surfaces of the coarse-grained samples.

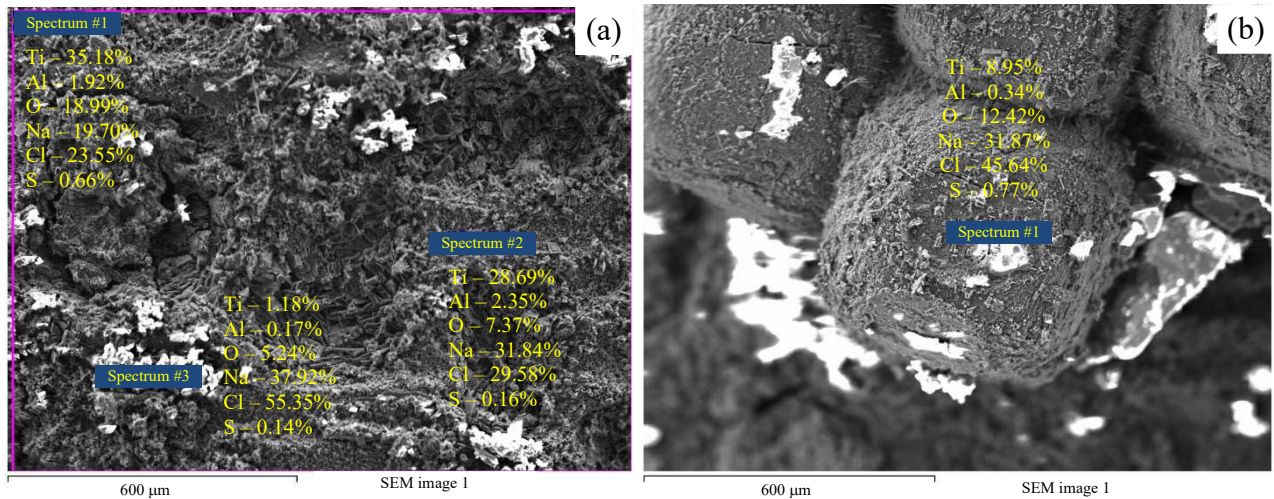

**Fig. 13**. EDS micro-analysis of the composition of salt deposits on the surfaces of coarse-grained Ti–2.5Al–2.6 alloy samples after the HSC tests. SEM.

Fig. 14 shows the typical appearance of corrosion defects on the surfaces of coarse-grained alloy samples after HSC testing. As evident from Fig. 14, a combination of pitting corrosion and IGC occurs in the coarse-grained titanium alloy, developing simultaneously. Notably, the largest IGC defects were often located at the bottom of the corrosion pits (Fig. 14a), suggesting that IGC defects formed initially during the HSC testing. Subsequently, the corrosion pits rapidly grew, engulfing the IGC defects. Additionally, the IGC defects on the sample surfaces were primarily localized in the areas where the α′-phase plates and β-phase particles were located along the grain boundaries. This results in the characteristic plate-like appearance of the IGC defects (Fig. 14b).

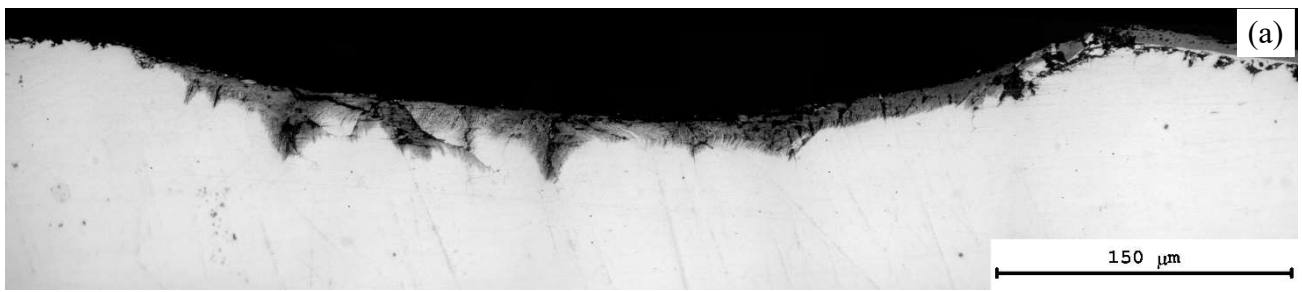

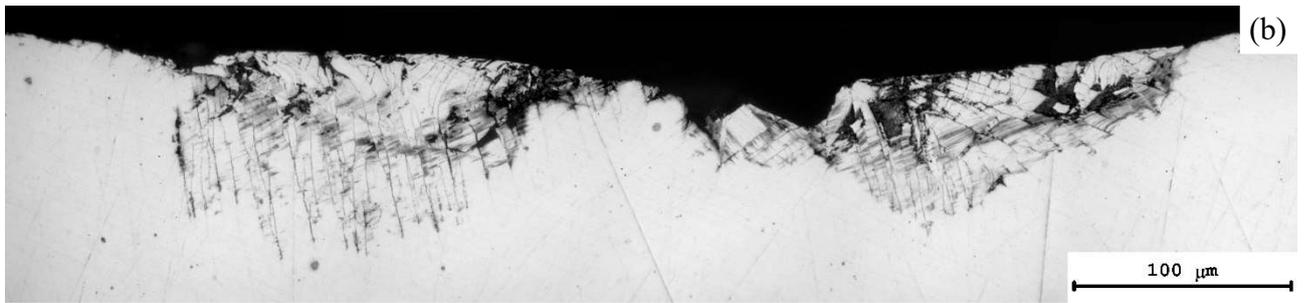

**Fig. 14**. Characteristics of corrosion defects on the surface of the coarse-grained Ti–2.5Al–2.6Zr alloy after the HSC test: (a) a cross-section view and (b) a longitudinal section view.

Two types of corrosion defects were observed on the surfaces of the UFG alloy samples after HSC testing: large corrosion pits with diameters of ~1 mm and depths up to 200 µm. A shallow IGC defects were noted at the bottoms of the corrosion pits (Fig. 15a). The characteristic size of the fragments, corresponding to the distances between the IGC defects, was ~10 µm. This result indicates the start of grain growth during the HSC testing. This conclusion agrees with the results of metallographic studies (Fig. 15b). It should be noted that the non-uniform macrostructure of the sample after HSC testing (Fig. 15c) inherits the non-uniformity of the original macrostructure of the titanium alloy after hot rolling (Fig. 4b). The sections of the original lightly deformed macrostructure in the Ti–2.5Al–2.6Zr alloy sample are indicated by dashed lines in Fig. 15c.

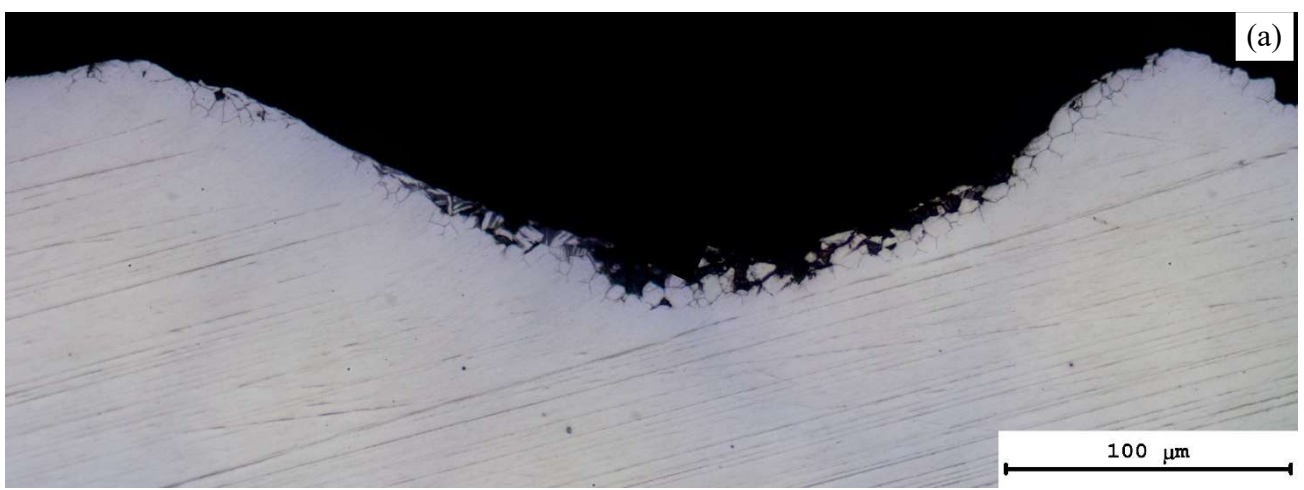

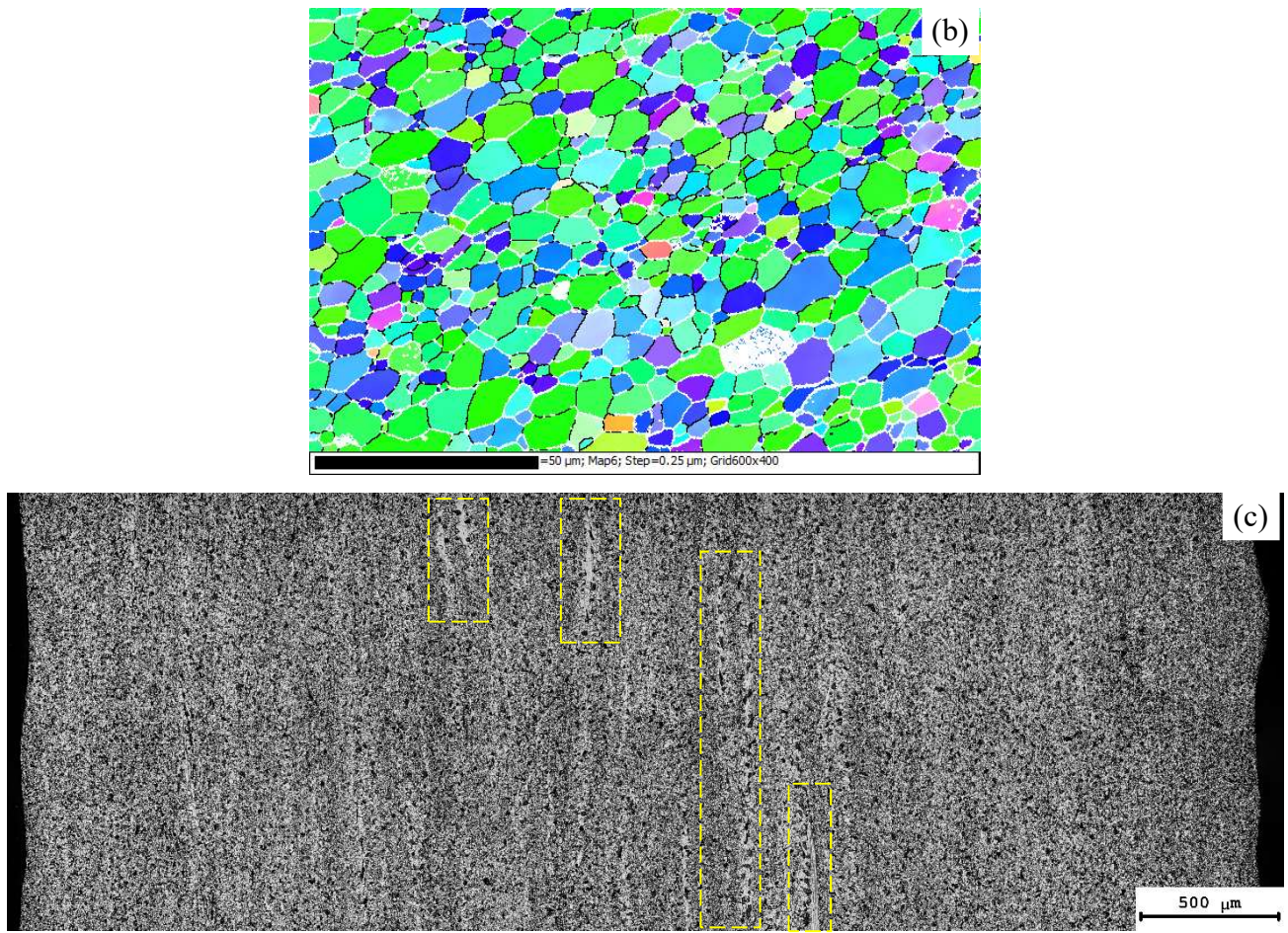

**Fig. 15**. Character of corrosion defects on the surface of the UFG alloy following the HSC test (a), along with the micro- (b) and macrostructure (c) of the UFG alloy after the HSC test. Fig. 15b displays a metallographic optical microscopy image of the longitudinal section of the sample.

Notably, the intensity of pitting corrosion in the Ti–2.5Al–2.6Zr UFG alloy was approximately 1.5 to 2 times greater than that in the coarse-grained alloy. Figs. 16a and 16b show that the reduction in the cross-sectional area of the UFG alloy samples was more significant than the respective value for the coarse-grained alloy. The growth rate of the corrosion pits in the UFG alloy is notably high, with the corrosion pits almost entirely engulfing the IGC defects. It is also important to mention that the nature of the IGC defects in the UFG alloy differs from those in the coarse-grained alloy. As demonstrated in Fig. 16a, typical IGC, induced by increased corrosion-aggressive impurities at the grain boundaries, was frequently observed in the UFG alloy (refer to [32-34]).

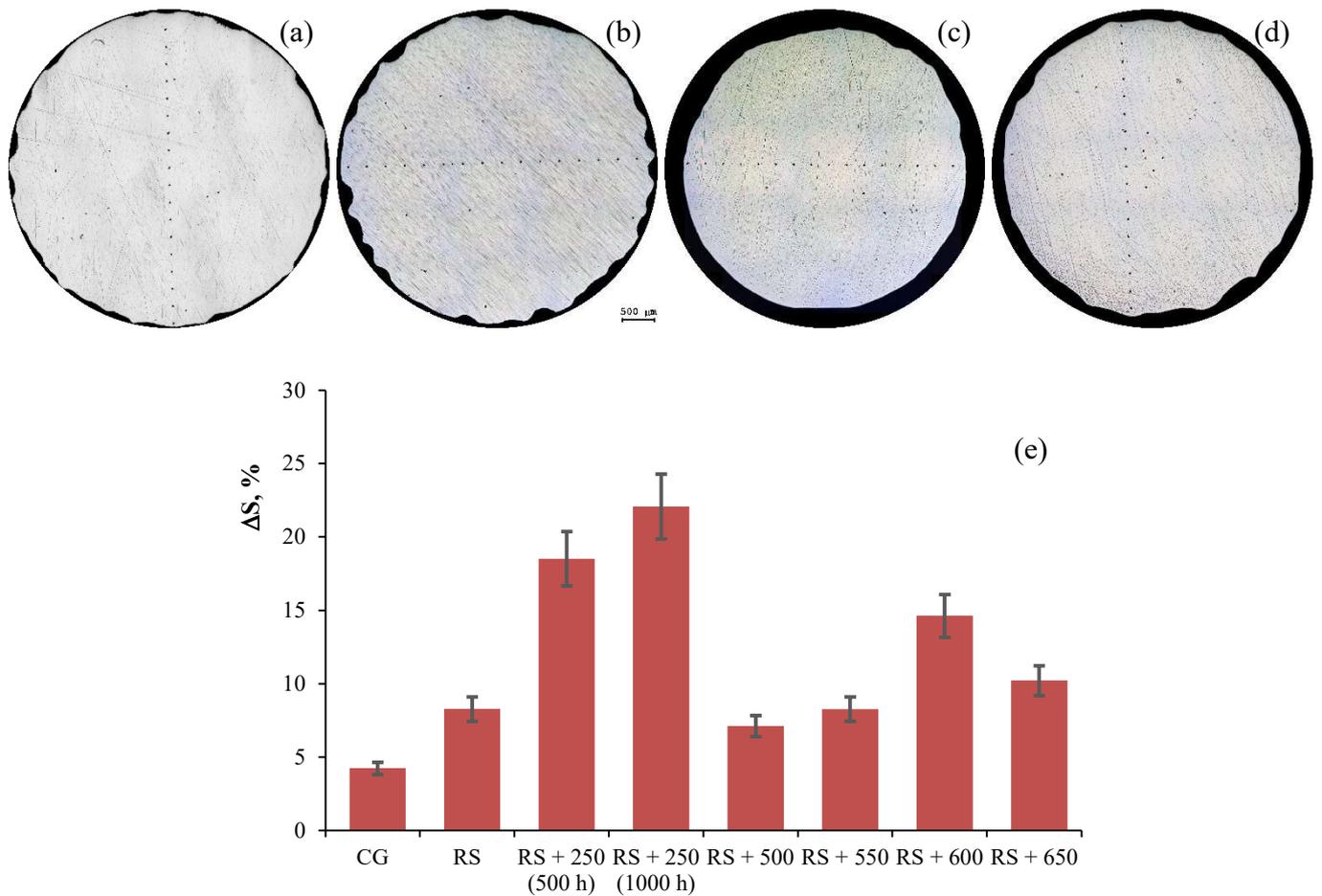

**Fig. 16.** Effect of annealing mode on the corrosion intensity of Ti–2.5Al–2.6Zr alloy. Cross sections of the samples post-HSC tests: (a) coarse-grained alloy; (b) alloy after RS; (c) alloy after annealing at 250°C for 1000 h; (d) alloy after annealing at 600°C for 30 min. To simplify the comparison, the samples were encased in bakelite moulds measuring 5 mm in diameter after testing. (e) Results from measuring changes in the cross-sectional areas (ΔS) of the samples after the HSC testing.

Annealing did not alter the composition of the HSC films, but it did influence the nature of the HSC process. The analysis of how annealing regimes affect the corrosion resistance of the UFG alloy revealed that increasing the annealing temperature leads to a rise in the overall corrosion rate. The presence of large corrosion pits on the surfaces of the annealed samples was fewer than those on the surfaces of both coarse-grained and UFG samples (Fig. 16). The depth of the IGC defects on the surfaces of the annealed samples did not exceed 30–40 μm (Fig. 17), which is 1.5–2 times greater than that observed in the unannealed alloy in its as-rolled state.

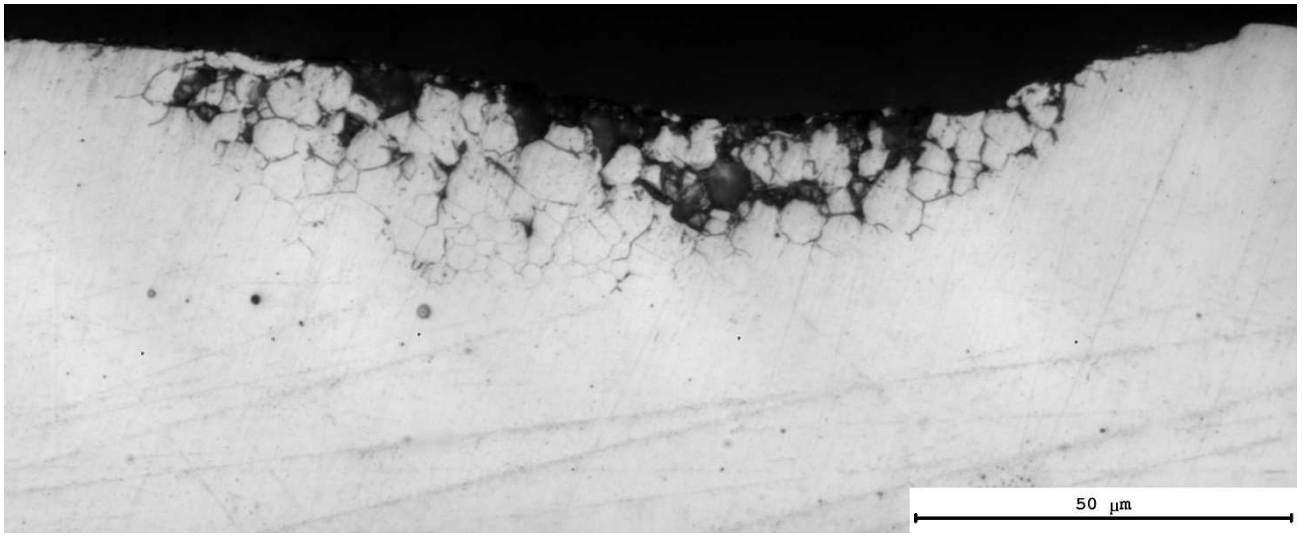

**Fig. 17** IGC on the surface of Ti–2.5Al–2.6Zr alloy after annealing at 650°C for 30 min.

The overall results of the corrosion tests show that the coarse-grained Ti–2.5Al–2.6Zr alloy exhibited increased resistance to pitting corrosion but demonstrated a tendency towards IGC. The heavily deformed titanium α–alloy after RS exhibited less resistance to pitting corrosion (Table 2). Increasing the annealing duration at 250°C and raising the annealing temperature for 30 min from 500°C to 650°C led to in an increase in the uniform corrosion rate and a minor increase in the IGC depth. A particularly unusual result is the significantly higher uniform corrosion rate of the Ti–2.5Al–2.6Zr alloy after prolonged annealing at 250°C (Fig. 16).

**Table 2**. Effect of annealing on the susceptibility of Ti–2.5Al–2.6Zr α-alloy: (−) indicates a negligible contribution, (+) signifies a small contribution, (++) denotes a moderate contribution and (+++) represents a significant contribution.

| Alloy state | IGC | Pitting corrosion | Uniform corrosion |
|---|---|---|---|
| Initial state | + | + | − |
| After RS | + | ++ | + |
| After RS and annealing at 250°C | − | + | +++ |
| After RS and annealing at 500–650°C | + | + | ++ |

## 4. Discussion

### 4.1 Effect of rotary swaging

The analysis of literature has shown that typically, α-titanium alloys demonstrate high resistance to HSC at limited air access in the testing chamber. However, the presence of air and water vapour into the testing chamber accelerates IGC in titanium α- and near-α alloys [8-11]. Under specific testing conditions (including salt composition, testing temperature, oxygen flow rate and air humidity), the susceptibility of α- and near-α titanium alloys to IGC depends primarily on the microstructure as well as the chemical and phase compositions of the grain boundaries.

Notably, the intergranular HSC rate in titanium alloys is complex, influenced by a series of chemical and electrochemical reactions [8]. The rates of the chemical reactions occurring at the grain boundaries depend on the concentrations of the alloying elements at the grain boundaries and the dissolution rate of β-phase particles located at these boundaries. The β-phase particles can disrupt the integrity of the oxide film on the surface of titanium samples, thereby contributing to additional acceleration of the chemical corrosion rate. It is also notable that the oxide films on Ti–Zr alloy surfaces start to form at the grain boundaries and spread across the grains [55]. Therefore, the thicknesses of the oxide films on the surfaces near the grain boundaries usually surpass those on the grain surfaces themselves [55]. The presence of β-phase particles reduces the thicknesses of the oxide films and hence their protective characteristics.

First, we analyse the effect of cold severe plastic deformation on the HSC resistance of the Ti–2.5Al–2.6Zr alloy.

In its initial state, the coarse-grained alloy exhibits increased susceptibility to IGC because of the presence of β-phase particles at the low-angle boundaries of the α′-grains (Fig. 3b). These β-phase particles contain a higher concentration of alloying elements, enhancing the stability of the titanium FCC lattice at room temperature [4]. Studies [56, 57] indicate that Zr contributes minimally to stabilising the β-phase in Ti–Al–V alloys. The β–phase of titanium is characterised by a more negative electrode potential ($E_β$) than the α–phase ($E_α$), leading to its intensive etching during corrosion tests. This results in a shift of the corrosion potential ($E_{corr}$) in the two-phase α+β alloys (such as Ti–6Al–4V, Ti–6Al–2Sn–4Zr–2Mo) towards more negative potential values compared to pure Ti [58]. In the Ti–2.5Al–2.6Zr alloy, the concentrations of alloying elements are sufficiently low, resulting in a low content of β-phase particles in the α–alloys is relatively small. The precipitation of β-phase particles leads to the formation of micro-galvanic couples within the structure of the α-alloy, which accelerates corrosion along the α/β interphase boundaries [59-61].

The second factor contributing to IGC in the Ti–2.5Al–2.6Zr α-alloy is the increased concentration of Zr at the HAGBs. Notably, the Zr concentration at the HAGBs is maintained even after the formation of the UFG microstructure through ECAP (Fig. 4g). As shown in Fig. 4g, the Zr concentration at the HAGBs is approximately twice that within the α-Ti grains. Clearly, the contribution of the HAGBs to the susceptibility of the titanium alloy to IGC will depend on the Zr concentration at the grain boundaries. If, for any reason, the local Zr concentration at the grain boundaries increases, IGC will be accelerated. However, it is important to note that the contribution of Zr-enriched HAGBs to IGC will be less than that of micro-galvanic couples at the interphase α/β boundaries.

During RS, the sizes of Ti grains and β-phase particles located along the grain boundaries are refined. Electron microscopy results do not reveal micron-sized β-phase particles along the grain boundaries in the UFG Ti alloy after ECAP (Figs. 4e and 4g). Thus, after ECAP, the UFG alloy exhibits increased resistance towards IGC because the contribution of the β–phase particles and interphase α/β–boundaries is considerably reduced. In addition, severe plastic deformation of the grains leading to an increase in the density of lattice dislocations decreases the value of the electrode potential of the α–phase [58, 62]. Therefore, the potential difference between the α– and β–phases ($\Delta E = E_\alpha - E_\beta$) will decrease, thereby reducing the contribution of the interphase α/β–boundaries to the IGC intensity of the Ti alloy.

The discussion focuses on the origin of the increased pitting corrosion rate observed in UFG alloys. It is typically assumed that refining of the microstructure leads to the formation of a thicker and more stable oxide film on the surfaces of Ti [63] and Ti–Zr alloys [55]. Therefore, the pronounced formation of corrosive pits in the UFG alloy emerges as a rather unexpected result. It is commonly assumed that the origin of pitting corrosion in Ti–Al–Zr alloys points to Ti–Fe intermetallics as a potential cause [53]. Nevertheless, it seems improbable that individual nanoparticles could instigate the formation of corrosive pits exceeding 100 μm in size (Fig. 15).

Notably, IGC and pitting corrosion processes in the Ti–2.5Al–2.6Zr alloy are related to each other. The results of metallographic investigations show that the IGC defects are located at the bottom of corrosive pits (Fig. 15). Additionally, it is important to note that accurately measuring the IGC depth in the UFG titanium alloy is practically impossible because the IGC defects are almost entirely captured by swiftly expanding corrosive pits.

The initiation of corrosive pits is chemically attributed to chloride ions on the surface of the protective Ti oxide film. During HSC testing, HCl is formed, further contributing to the degradation of the protective oxide film on the Ti surface. Such corrosive pit initiations primarily occur on the surfaces of Ti–Fe particles (Figs. 6b and 6g, and see [53]). The authors of [53] suggested that the increase in the corrosion rate is primarily associated with the growth of TiFe and $TiFe_2$ particles, which were originally present in the alloy. According to the model proposed in [53], these particles may affect the composition of the oxide films on the Ti alloy surface and facilitate the diffusion of oxygen into the Ti surface and the accumulation of hydrogen in the Ti alloy.

In the coarse-grained Ti alloy, a high IGC rate within the lamellar microstructure regions is the origin of accelerated pit growth (Fig. 14b). The presence of α′-phase plates alternating with equiaxial α-phase grains contributes to the localisation of IGC areas on the surface of the coarse-grained alloy (Fig. 14b). Rapid corrosive destruction of the α′-phase areas leads to the emergence of large corrosive pits on the surface of the coarse-grained alloy.

During RS, the α- and α′-grains are refined, resulting in a swirling nonuniform macrostructure (Fig. 4a). Within this nonuniform macrostructure, areas of weakly deformed coarse-grained microstructure alternate with areas of severely deformed UFG microstructure. This microstructure pattern is characteristic of low degrees of deformation during RS [41]. The accelerated destruction of the heavily deformed regions within the nonuniform macrostructure, which remains in the sample structure even after prolonged HSC testing (Fig. 15b), may contribute to the formation of deep corrosion pits.

### 4.2 Low-temperature annealing at 250°C

The analysis of the experimental results indicates that prolonged (500–1000 h) low-temperature annealing at 250°C results in a notable increase in the general corrosion rate of the UFG alloy (Fig. 16 and Table 2).

The high corrosion resistance of the Ti–Al–Zr alloy is primarily due to the positive effect of Zr on the corrosion resistance of Ti and the stability of the oxide film on the surface of the titanium alloy [56, 57, 64, 65]. The detrimental effect of Zr on the corrosion resistance of Ti alloys becomes significant only when Zr concentrations exceed 30% [55, 66]. It is widely believed that Zr atoms enhance the hardness and corrosion resistance of the Ti oxide by integrating into it. With higher concentrations of Zr, particles of Zr oxide form, exhibiting robust resistance to chemical and electrochemical corrosion [57]. Minor additions of Zr (up to 5 wt.%) improve the open-circuit potential and elevate the corrosion potential [64]. This reduces the corrosion current $i_{corr}$ [64, 67] and boosts resistance of Ti to oxidation [65].

Analysis of the microstructure investigation results reveals that after annealing for 30 min at 400°C, Zr particles precipitate in the UFG alloy (Fig. 6a) along with Ti–Fe intermetallics (Figs. 6c and 6g). Following a 30-min annealing at 600°C, the Zr particles completely dissolve inside the recrystallized Ti grains. The precipitation of Zr particles leads to a decrease in Zr concentration in the crystal lattice, reducing the protective properties of the oxide films on the Ti alloy surface. The formation of Ti–Fe particles contributes to the accelerated degradation of the protective film composed of depleted Zr–Ti oxide $(Ti, Zr)_xO_y$. Therefore, the simultaneous formation of Zr and Ti–Fe particles accelerates the general corrosion of the Ti–2.5Al–2.6Zr alloy, as demonstrated in the experiment (Fig. 16 and Table 2).

Let us estimate the characteristic diffusion time ($\tau_{diff}$) of Zr in Ti at 400°C and 250°C. To calculate $\tau_{diff}$, the equation $\tau_{diff} = x^2/D_{Zr}$ is used, where $x$ is the characteristic diffusion path of Zr in the α-Ti lattice, $D_{Zr} = D_{0Zr}\exp(-Q_{Zr}/kT)$ is the diffusion coefficient of Zr in the α-Ti lattice, $D_{0Zr} = 4 \times 10^{-3}$ m$^2$/s [68] is the pre-exponential factor in the equation for the diffusion coefficient, $Q_{Zr}$ = 304 kJ/mol [68] ~ 18.8 $kT_m$ is the activation energy of Zr diffusion in the α-Ti lattice, $T_m$ = 1943 K is the melting temperature of Ti and $k$ is the Boltzmann constant. Assuming $x$ at $T_1$ = 400°C ($\tau_1$ = 30 min) to be $x_1 = d_{1/2} \approx 0.5$ μm and $d_{2/2} \approx 5$ μm at $T_2$ = 250°C, one obtains $\tau_1/\tau_2 \approx 1.2$. Therefore, the diffusion-controlled processes occurring at 400°C for ~30 min will take ~1000 h at 250°C. It is also important to note that in the UFG materials, the rates of many non-equilibrium diffusion processes may be limited by the grain boundary diffusion, the activation energy of which in Ti is anomalously low compared to other metals [69].

### 4.3 Annealing at 500–650°C

The analysis of the obtained results indicates that an elevation in annealing temperature increases the uniform corrosion rate (Fig. 16 and Table 2). These findings are consistent with those of [53], which explored the corrosion resistance of the coarse-grained Ti–2.25Al–2.24Zr alloy in an aqueous ammonia solution (pH 9.98, 360°C, for 220 days). The alloy samples, with an average grain size of 20 μm, were annealed at 500°C, 600°C, 700°C and 800°C. It was demonstrated that an increase in the annealing temperature led to a rise in the mass gain of the samples, indicating an increased corrosion rate.

It should also be noted that, on average, the HSC rate after high-temperature annealing at 500–650°C is lower than after low-temperature annealing at 250°C (Fig. 16). In our view, the primary cause of pitting corrosion at elevated temperatures is the presence of Ti–Fe particles, the sizes of which may increase with the annealing temperature [53]. Additionally, we believe that the reduction in the uniform corrosion rate during high-temperature annealing results from the dissolution of Zr particles in the crystalline lattice of the Ti–2.5Al–2.6Zr alloy. This leads to an increased concentration of Zr incorporated into the protective Ti oxide film, thereby enhancing its corrosion resistance.

5. Conclusions

1. The HSC rate of the Ti–2.5Al–2.6Zr α-alloy (Russian industrial alloy PT-7M) at 250°C is influenced by the IGC intensity, pitting corrosion and uniform corrosion. The extent to which each type of corrosion contributes to the HSC rate is determined by the material microstructure under identical testing conditions. The presence of β-phase particles and the concentration of Zr at the grain boundaries predominantly affect the IGC rate in the Ti alloy. The rate of pitting corrosion is dependent on the IGC rate, the presence of Ti–Fe particles and the characteristics of the heterogeneous macrostructure resulting from RS. The rate of uniform corrosion, evidenced by mass loss and reduction in the cross-sectional area of specimens, is primarily influenced by the protective properties of the oxide film, which are contingent on the concentration of Zr within it.

2. Cold RS results in the refinement of the β-phase particles and a nonuniform macrostructure, wherein regions of heavily deformed metal with UFG microstructures alternate with regions of weakly deformed metal. These weakly and heavily deformed metal regions display varying HSC rates, leading to a reduction in the IGC rate and an increase in the pitting corrosion rate.

3. Low-temperature annealing results in the precipitation of Zr particles and Ti–Fe intermetallics (TiFe and TiFe$_2$). This diminishes the protective qualities of the oxide films on the Ti surfaces, thereby increasing the uniform corrosion rate. The uniform corrosion rate in annealed Ti alloys surpasses the rates of pitting corrosion and IGC when tested in crystalline NaCl at 250°C.

**CRediT Authorship Contribution Statement:** V.N. Chuvil'deev - Formal analysis, Methodology, Writing - review and editing, Project administration, Funding acquisition, Supervision, Resources; A.A. Murashov – Investigation (SEM), C.V. Likhnitskii – Investigation (Metallography, Microhardness), N.V. Melekhin – Investigation (Compression tests), K.A.

Rubtsova – Investigation (XRD analysis), N.Yu. Tabachkova – Investigation (TEM), A.V. Nokhrin & A.I. Malkin – Formal analysis, Writing - original draft preparation, Data curation, Visualization; A.M. Bakhmetyev, P.V. Tryaev, R.A. Vlasov – Investigation (HSC test).

**Funding:** The work was carried out within the framework of the grant #H-498-99_2021-2023 (#075-15-2021-1332) of the Federal Academic Leadership Priority Program Priority-2030 of the Ministry of Science and Higher Education of the Russian Federation and the grant #075-03-2023-096 (FSWR-2023-0037) of the Ministry of Science and Higher Education of the Russian Federation.

**Acknowledgments:** TEM investigations of the microstructure was carried out using the equipment of the Center Collective Use "Materials Science and Metallurgy" of the National University of Science and Technology "MISIS" (Moscow, Russia). The authors thank to D.N. Kotkov (UNN, Nizhny Novgorod, Russia) and D.A. Zotov (UNN, Nizhny Novgorod, Russia) for UFG titanium alloys obtained via cold Rotary Swaging.

**Conflicts of Interest:** The authors declare that they have no known competing financial interests or personal relationships that could have appeared to influence the work reported in this paper.

**Data Availability:** Data will be available on the request.